\documentclass[prd,twocolumn,eqsecnum,superscriptaddress,showpacs,letterpaper,altaffilletter,amssymb,amsmath,amsfonts,aps]{revtex4}

\makeatletter
\def\@fnsymbol#1{\ifcase#1\or * \or  $+$ \or  \$ \or \#  \or \dag \or \ddag \or
$\mathsection$ \or $ \mathparagraph$ \or $\|$  \or
\textordfeminine \or \textbullet   \or ** \or $++$ \or  \$\$ \or
\#\#  \or \dag\dag \or \ddag\ddag \or $\mathsection\mathsection$
\or $ \mathparagraph\mathparagraph$ \or $\|\|$  \or
\textordfeminine\textordfeminine \or \textbullet \textbullet \or
*** \or $+++$ \or  \$\$\$ \or \#\#  \or \dag\dag \or \ddag\ddag
\or $\mathsection \mathsection\mathsection$ \or $ \mathparagraph
\mathparagraph\mathparagraph$ \or $\|\|\|$  \or
\textordfeminine\textordfeminine\textordfeminine \or
\textbullet\textbullet\textbullet \or \else \@ctrerr\fi}
\makeatother

\usepackage{color}
\usepackage{times}
\usepackage{graphicx}
\usepackage{fancyhdr}
\usepackage{float}

\def\thercsid{\relax}

\renewcommand{\today}{\number\day\space\ifcase\month\or
  January\or February\or March\or April\or May\or June\or
  July\or August\or September\or October\or November\or December\fi
  \space\number\year}

\def\be{\begin{equation}}
\def\ee{\end{equation}}
\def\bi{\begin{itemize}}
\def\ei{\end{itemize}}
\def\ben{\begin{enumerate}}
\def\een{\end{enumerate}}

\newcommand\ligodoc{P040007-08-D}

\begin{document}

\title{
A Search for Gravitational Waves Associated with the Gamma Ray
Burst GRB030329 Using the LIGO Detectors
\\
{\color{blue} {\large LIGO-\ligodoc} }
\\
}

\begin{abstract}
\vspace*{0.2in}
We have performed a search for bursts of gravitational waves associated with the very bright Gamma Ray Burst GRB030329, using the two detectors at the LIGO Hanford Observatory. Our search covered the most sensitive frequency range of the LIGO detectors (approximately 80-2048~Hz), and we specifically targeted signals shorter than $\simeq$150~ms. Our search algorithm looks for excess correlated power between the two interferometers and thus makes minimal assumptions about the gravitational waveform. We observed no candidates with gravitational wave signal strength larger than a pre-determined threshold. We report frequency dependent upper limits on the strength of the gravitational waves associated with GRB030329. Near the most sensitive frequency region, around $\simeq$250~Hz, our root-sum-square (RSS) gravitational wave strain sensitivity for optimally polarized bursts was better than h$_{RSS}{\simeq}$6$\times$10$^{-21}$~Hz$^{-1/2}$. Our result is comparable to the best published results searching for association between gravitational waves and GRBs.
\end{abstract}


\date[\relax]{ RCS \thercsid; compiled \today }


\newcommand*{\AG}{Albert-Einstein-Institut, Max-Planck-Institut f\"ur Gravitationsphysik, D-14476 Golm, Germany}

\affiliation{\AG}

\newcommand*{\AH}{Albert-Einstein-Institut, Max-Planck-Institut f\"ur Gravitationsphysik, D-30167 Hannover, Germany}

\affiliation{\AH}

\newcommand*{\AN}{Australian National University, Canberra, 0200, Australia}

\affiliation{\AN}

\newcommand*{\CH}{California Institute of Technology, Pasadena, CA  91125, USA}

\affiliation{\CH}

\newcommand*{\DO}{California State University Dominguez Hills, Carson, CA  90747, USA}

\affiliation{\DO}

\newcommand*{\CA}{Caltech-CaRT, Pasadena, CA  91125, USA}

\affiliation{\CA}

\newcommand*{\CU}{Cardiff University, Cardiff, CF2 3YB, United Kingdom}

\affiliation{\CU}

\newcommand*{\CL}{Carleton College, Northfield, MN  55057, USA}

\affiliation{\CL}

\newcommand*{\FN}{Fermi National Accelerator Laboratory, Batavia, IL  60510, USA}

\affiliation{\FN}

\newcommand*{\HC}{Hobart and William Smith Colleges, Geneva, NY  14456, USA}

\affiliation{\HC}

\newcommand*{\IU}{Inter-University Centre for Astronomy  and Astrophysics, Pune - 411007, India}

\affiliation{\IU}

\newcommand*{\CT}{LIGO - California Institute of Technology, Pasadena, CA  91125, USA}

\affiliation{\CT}

\newcommand*{\LM}{LIGO - Massachusetts Institute of Technology, Cambridge, MA 02139, USA}

\affiliation{\LM}

\newcommand*{\LO}{LIGO Hanford Observatory, Richland, WA  99352, USA}

\affiliation{\LO}

\newcommand*{\LV}{LIGO Livingston Observatory, Livingston, LA  70754, USA}

\affiliation{\LV}

\newcommand*{\LU}{Louisiana State University, Baton Rouge, LA  70803, USA}

\affiliation{\LU}

\newcommand*{\LE}{Louisiana Tech University, Ruston, LA  71272, USA}

\affiliation{\LE}

\newcommand*{\LL}{Loyola University, New Orleans, LA 70118, USA}

\affiliation{\LL}

\newcommand*{\MP}{Max Planck Institut f\"ur Quantenoptik, D-85748, Garching, Germany}

\affiliation{\MP}

\newcommand*{\MS}{Moscow State University, Moscow, 119992, Russia}

\affiliation{\MS}

\newcommand*{\ND}{NASA/Goddard Space Flight Center, Greenbelt, MD  20771, USA}

\affiliation{\ND}

\newcommand*{\NA}{National Astronomical Observatory of Japan, Tokyo  181-8588, Japan}

\affiliation{\NA}

\newcommand*{\NO}{Northwestern University, Evanston, IL  60208, USA}

\affiliation{\NO}

\newcommand*{\SC}{Salish Kootenai College, Pablo, MT  59855, USA}

\affiliation{\SC}

\newcommand*{\SE}{Southeastern Louisiana University, Hammond, LA  70402, USA}

\affiliation{\SE}

\newcommand*{\SA}{Stanford University, Stanford, CA  94305, USA}

\affiliation{\SA}

\newcommand*{\SR}{Syracuse University, Syracuse, NY  13244, USA}

\affiliation{\SR}

\newcommand*{\PU}{The Pennsylvania State University, University Park, PA  16802, USA}

\affiliation{\PU}

\newcommand*{\TC}{The University of Texas at Brownsville and Texas Southmost College, Brownsville, TX  78520, USA}

\affiliation{\TC}

\newcommand*{\TR}{Trinity University, San Antonio, TX  78212, USA}

\affiliation{\TR}

\newcommand*{\HU}{Universit{\"a}t Hannover, D-30167 Hannover, Germany}

\affiliation{\HU}

\newcommand*{\BB}{Universitat de les Illes Balears, E-07122 Palma de Mallorca, Spain}

\affiliation{\BB}

\newcommand*{\BR}{University of Birmingham, Birmingham, B15 2TT, United Kingdom}

\affiliation{\BR}

\newcommand*{\FA}{University of Florida, Gainesville, FL  32611, USA}

\affiliation{\FA}

\newcommand*{\GU}{University of Glasgow, Glasgow, G12 8QQ, United Kingdom}

\affiliation{\GU}

\newcommand*{\MU}{University of Michigan, Ann Arbor, MI  48109, USA}

\affiliation{\MU}

\newcommand*{\OU}{University of Oregon, Eugene, OR  97403, USA}

\affiliation{\OU}

\newcommand*{\RO}{University of Rochester, Rochester, NY  14627, USA}

\affiliation{\RO}

\newcommand*{\UW}{University of Wisconsin-Milwaukee, Milwaukee, WI  53201, USA}

\affiliation{\UW}

\newcommand*{\WU}{Washington State University, Pullman, WA 99164, USA}

\affiliation{\WU}

\author{B.~Abbott}    \affiliation{\CT}

\author{R.~Abbott}    \affiliation{\LV}

\author{R.~Adhikari}    \affiliation{\LM}

\author{A.~Ageev}    \affiliation{\MS}  \affiliation{\SR}

\author{B.~Allen}    \affiliation{\UW}

\author{R.~Amin}    \affiliation{\FA}

\author{S.~B.~Anderson}    \affiliation{\CT}

\author{W.~G.~Anderson}    \affiliation{\TC}

\author{M.~Araya}    \affiliation{\CT}

\author{H.~Armandula}    \affiliation{\CT}

% ...

 \collaboration{The LIGO Scientific Collaboration, http://www.ligo.org}

 \noaffiliation

%

%

\maketitle

\section{Introduction}\label{sec:Introduction}

Gamma Ray Bursts (GRBs) are short but very energetic pulses of gamma
rays from astrophysical sources, with duration ranging between 10~ms
and 100~s. GRBs are historically divided into two
classes~\cite{Kouveliotou93,Meszaros03} based on their duration:
``short" ($<$~2~s) and ``long" ($>$~2~s). Both classes are
isotropically distributed and their detection rate can be as large
as one event per day. The present consensus is that long
GRBs~\cite{Meszaros03} are the result of the core collapse of
massive stars resulting in black hole formation. The violent
formation of black holes has long been proposed as a potential
source of gravitational waves. Therefore, we have reason to expect
strong association between GRBs and gravitational
waves~\cite{Fryer01,Davies02,Putten04}. In this paper, we report on
a search for a possible short burst of gravitational waves
associated with GRB030329 using data collected by the Laser
Interferometer Gravitational Wave Observatory (LIGO).

On March 29, 2003, instrumentation aboard the HETE-2
satellite~\cite{HETE04} detected a very bright GRB, designated
GRB030329. The GRB was followed by a bright and well-measured
afterglow from which a redshift~\cite{Price03} of $z=$0.1685
(distance~$\simeq$800~Mpc~\cite{Matheson03}) was determined. After
approximately 10 days, the afterglow faded to reveal an underlying
supernova (SN) spectrum, SN2003dh~\cite{Sokolov03}. This GRB is the
best studied to date, and confirms the link between long GRBs and
supernovae.

At the time of GRB030329, LIGO was engaged in a 2-month long data
run. The LIGO detector array consists of three interferometers, two
at the Hanford, WA site and one at the Livingston, LA site.
Unfortunately, the Livingston interferometer was not operating at
the time of the GRB; therefore, the results presented here are based
on the data from only the two Hanford interferometers. The LIGO
detectors are still undergoing commissioning, but at the time of
GRB030329, their sensitivity over the frequency band 80 to 2048~Hz
exceeded that of any previous gravitational wave search, with the
lowest strain noise of ${\simeq}$6$\times$10$^{-22}$~Hz$^{-1/2}$
around 250~Hz.

A number of long GRBs have been associated with X-ray, radio and/or
optical afterglows, and the cosmological origin of the host galaxies
of their afterglows has been unambiguously established by their
observed redshifts, which are of order unity~\cite{Meszaros03}. The
smallest observed redshift of an optical afterglow associated with a
detected GRB (GRB980425~\cite{Kulkarni98,Iwamoto98,Galama98}) is
$z$=0.0085 ($\simeq$35~Mpc). GRB emissions are very likely strongly
beamed~\cite{Frail01,Putten03}, a factor that affects estimates of
the energy released in gamma rays (a few times 10$^{50}$ erg), and
their local true event rate (about 1 per year within a distance of
100Mpc).

In this search, we have chosen to look for a burst of gravitational
waves in a model independent way. Core collapse~\cite{Davies02},
black hole formation~\cite{Putten04,Bulik04} and black hole
ringdown~\cite{Hughes98,Jolien99} may each produce gravitational
wave emissions, but there are no accurate or comprehensive
predictions describing the gravitational wave signals that might be
associated with GRB type sources. Thus, a traditional matched
filtering approach~\cite{Helstrom68,InspiralS1} is not possible in
this case. To circumvent the uncertainties in the waveforms, our
algorithm does not presume any detailed knowledge of the
gravitational waveform and we only apply general bounds on the
waveform parameters. Based on current theoretical considerations, we
anticipate the signals in our detectors to be weak, comparable to or
less than the detector's noise~\cite{Muller2004,Fryer2004,Ott2004}.

This paper is organized as follows: Section II summarizes the
currently favored theories of GRBs and their consequences for
gravitational wave detection. Section III provides observational
details pertinent to GRB030329. Section IV briefly describes the
LIGO detectors and their data. Section V discusses the method of
analysis of the LIGO data. In Section VI we compare the events in
the signal region with expectations and we use simulated signal
waveforms to determine detection efficiencies. We also present and
interpret the results in this section. Section VII offers a
comparison with previous analyses, a conclusion, and an outlook
for future searches of this type.

\section{Production of gravitational waves in massive core collapses}\label{sec:Production_of_GW}

The apparent spatial association of GRB afterglows with spiral arms,
and by implication star formation regions in remote galaxies, has
lead to the current ``collapsar'' or ``hypernova'' scenario
~\cite{MacFadyen99,Heger2003} in which the collapse of a rotating,
massive star to a Kerr black hole can lead to relativistic ejecta
emitted along a rotation axis and the associated production of a GRB
jet. The identification of GRB030329 with the supernova SN2003dh
(section 3 below) gives further support to this association. This
observation is consistent with the theory that the GRB itself is
produced by an ultra-relativistic jet associated with a central
black hole. Stellar mass black holes in supernovae must come from
more massive stars. Ref. ~\cite{Heger2003} presents ``maps'' in the
metallicity-progenitor mass plane of the end-states of stellar
evolution and shows that progenitors with 25 $M_{\odot}$ can produce
black holes by fall-back accretion.

The observed pulsar kick velocities  of  $\simeq$500~km/s hint at a
strong asymmetry around the time of maximum compression, which may
indicate deviations from spherical symmetry in the progenitor. The
resulting back reaction on the core from the neutrino heating
provides yet another potential physical mechanism for generating a
gravitational wave signal. In the model of ~\cite{Burrows1996} it
imparts a kick of 400-600 km/s and an induced gravitational wave
strain roughly an order of magnitude larger than in
~\cite{Muller2004} and an order of magnitude smaller than
~\cite{Dimmelmeier2002}.

Theoretical work on gravitational wave (GW) signals in the process
of core-collapse in massive stars has advanced much in recent years,
but still does not provide detailed waveforms. Current models take
advantage of the increase in computational power and more
sophisticated input physics to include both 2D and 3D calculations,
utilizing realistic pre-collapse core models and a detailed, complex
equation of state of supernovae that produce neutron stars. The most
recent studies by independent groups give predictions for the strain
amplitude within a similar range, despite the fact that the dominant
physical mechanisms for gravitational wave emission in these studies
are different ~\cite{Muller2004, Fryer2004, Fryer2002, Zwerger1997,
Dimmelmeier2002}. The calculations of ~\cite{Muller2004} are
qualitatively different from previous core collapse simulations in
that the dominant contribution to the gravitational wave signal is
neutrino-driven convection, about 20 times larger than the
axisymmetric core bounce gravitational wave signal.

The applicability of the above models to GRBs is not clear, since
the model endpoints are generally neutron stars, rather than black
holes. Another recent model involves accretion disks around Kerr
black holes~\cite{vanPutten2003}, subject to non-axisymmetric
Papaloizou-Pringle instabilities~\cite{Papaloizou1984} in which an
acoustic wave propagates toroidally within the fallback material.
They are very interesting since they predict much higher
amplitudes for the gravitational wave emission.

For our search, the main conclusion to draw is that in spite of the
dramatic improvement in the theoretical models, there are no
gravitational waveforms that could be reliably used as templates for
a matched filter search, and that any search for gravitational waves
should ideally be as waveform independent as practical. Conversely,
detection of gravitational waves associated with a GRB would almost
certainly provide crucial new input for GRB/SN astrophysics. It is
also clear that the predictions of gravitational wave amplitudes are
uncertain by several orders of magnitude, making it difficult to
predict the probability to observe the gravitational wave signature
of distant GRBs.

The timeliness of searching for a gravitational wave signal
associated with GRBs is keen in light of the recent work by
~\cite{Fryer2004} and \cite{Muller2004}. ~\cite{Muller2004} finds
that the signal due to neutrino convection exceeds that due to the
core bounce and therefore a chaotic signal would be expected.
Studies with simplified or no neutrino transport (e.g.,
~\cite{Fryer2004}, \cite{Ott2004}) find the core-bounce to be the
dominant contributor to the GW signal. The large-scale, coherent
mass motions involved in the core bounce leads to a predicted
gravitational wave signal resembling a damped sinusoid.

\section{GRB030329 RELATED OBSERVATIONAL RESULTS }\label{sec:Observations}

\subsection{Discovery of GRB 030329 and its afterglow}\label{sec:Afterglow}

\begin{figure}[!t]
\includegraphics[angle=0,width=84mm]{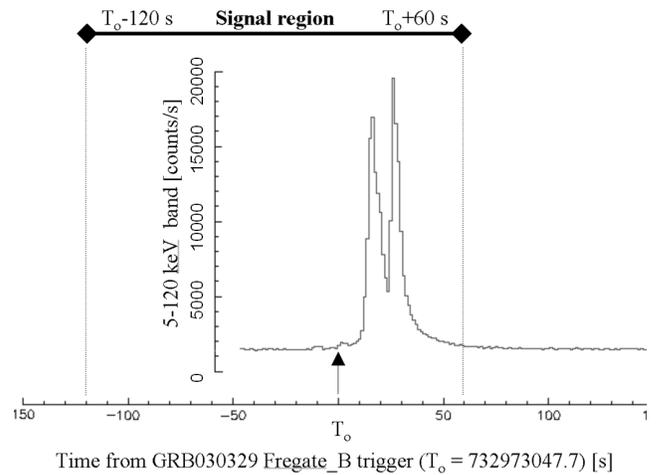}
\caption{The GRB030329 light curve as measured by the HETE-2 FREGATE
B instrument. The arrow indicates the HETE trigger time. The signal
region analyzed in this study is indicated by the horizontal bar at
the top. This figure is the courtesy of the HETE collaboration. }
\label{fig:HETE}
\end{figure}

On March 29, 2003 at 11:37:14.67 UTC, a GRB triggered the FREGATE
instrument on board the HETE-2
satellite~\cite{HETE04,Vanderspek03,Ricker03,Vander2004}. The GRB
had an effective duration of $\simeq$50~s, and a fluence of
1.08$\times$10$^{-4}$~erg/cm$^{2}$ in the 30-400~keV
band~\cite{Vander2004}. The KONUS detector on board the Wind
satellite also detected it~\cite{Golenetskii03}, triggering about
15~seconds after HETE-2. KONUS observed the GRB for about 35
seconds, and measured a fluence of
1.6$\times$10$^{-4}$~erg~/~cm$^{2}$ in the 15-5000~keV band. The
measured gamma ray fluences place this burst among the brightest
GRBs. Figure~\ref{fig:HETE} shows the HETE-2 light curve for
GRB030329~\cite{HETE030329}.

The rapid localization of the GRB by HETE ground analysis gave an
accurate position which was distributed about 73 minutes after the
original trigger. A few hours later, an optical
afterglow~\cite{Lipkin03,Price03} was discovered with magnitude
R=12.4, making it the brightest optical counterpart to any GRB
detected to date. The RXTE~\cite{Bradt93} satellite measured a X-ray
flux of 1.4$\times$10$^{-10}$~erg~s$^{-1}$~cm$^{-2}$ in the 2-10 keV
band about 4h51m after the HETE trigger, making this one of the
brightest X-ray afterglows detected by RXTE~\cite{Marshall03}. The
National Radio Astronomy Observatory (NRAO) observed ~\cite{GCN2014}
the radio afterglow, which was the brightest radio afterglow
detected to date~\cite{Berger03}. Spectroscopic measurements of the
bright optical afterglow~\cite{Greiner03} revealed emission and
absorption lines, and the inferred redshift ($z$~=~0.1685,
luminosity distance $D_L~\cong~$800~Mpc) made this the second
nearest GRB with a measured distance. To date, no host galaxy has
been identified. It is likely that numerous other GRBs have been
closer than GRB030329, but the lack of identified optical
counterparts has left their distances undetermined.

Spectroscopic measurements~\cite{Stanek03,Mazzali03,Matheson03},
about a week after the GRB trigger, revealed evidence of a
supernova spectrum emerging from the light of the bright optical
afterglow, which was designated SN2003dh. The emerging supernova
spectrum was similar to the spectrum of SN1998bw a week before its
brightness maximum~\cite{Hjorth03,Greiner03N}.

SN1998bw was a supernova that has been spatially and temporally
associated with GRB980425~\cite{Kulkarni98,Iwamoto98,Galama98}, and
was located in a spiral arm of the barred spiral galaxy ESO 184-G82
at a redshift of $z~=$~0.0085 ($\simeq$35~Mpc), making it the
nearest GRB with a measured distance. The observed spectra of
SN2003dh and SN1998bw, with their lack of hydrogen and helium
features, place them in the Type Ic supernova class. These
observations, together with the observations linking GRB980425
(which had a duration of ${\simeq}$23~s) to SN1998bw, make the case
that collapsars are progenitors for long GRBs more convincing. In
the case of SN1998bw, Woosley \emph{et al.}~\cite{Woosley99} and
Iwamoto \emph{et al.}~\cite{Iwamoto98} found that its observed
optical properties can be well modeled by the core collapse of a C+O
core of mass 6~M$_{\odot}$ (main sequence mass of 25~M$_{\odot}$)
with a kinetic energy of ${\simeq}$2$\times$10$^{52}$~ergs. This
energy release is about an order of magnitude larger than the
energies associated with typical supernovae.

\subsection{GRB030329 energetics}\label{sec:Energetics}

A widely used albeit naive quantity to describe the energy emitted
by GRBs is the total isotropic equivalent energy in gamma rays:
\begin{equation}\label{IEE}
E_{iso} = 4 \pi  (BC) D_{L}^{2} f / (1+z) \approx
2\times10^{52}~erg~.~
\end{equation}
where $f$ is the measured fluence in the HETE-2 waveband and BC is
the approximate bolometric correction for HETE-2 for long GRBs.
Using a ``Band spectrum"~\cite{Band93} with a single power law to
model the gamma ray spectrum, and using a spectral index,
$\beta=-$2.5, gives that the GRB's total energy integrated from 1
keV to 5 GeV is greater than that present in the band 30-400 keV by
a factor 2.2.

However, it is generally believed that GRBs are strongly beamed, and
that the change in slope in the afterglow light curve corresponds to
the time when enough deceleration has occurred so that relativistic
beaming is diminished to the point at which we ``see" the edge of
the jet. This occurs during the time in which the relativistic
ejecta associated with the GRB plows through the interstellar
medium, and the beaming factor $\Gamma^{-1}$, where $\Gamma$ is the
bulk Lorentz factor of the flow, increases from a value smaller than
the beaming angle $\Theta_j$, to a value larger than $\Theta_j$.
Effectively, prior to this time the relativistic ejecta appears to
be part of a spherical expansion, the edges of which cannot be seen
because they are outside of the beam, while after this time the
observer perceives a jet of finite width.

This leads to a faster decline in the light curves. Zeh \emph{et
al.} and Li \emph{et al.}~\cite{Zeh03,Li03} show that the initial
``break" or strong steepness in the light curve occurs at about 10
hours after the initial HETE-2 detection.

Frail \emph{et al.}~\cite{Frail01} give a parametric relation
between beaming angle $\Theta_{j}$, break time t$_{j}$, and
E$_{iso}$ as:
\begin{eqnarray}\label{Theta}
\Theta_{j} \approx 0.057 ~{\left(\frac{t_{j}}{24
hours}\right)}^{3/8} {\left(\frac{1+z}{2}\right)}^{-3/8}\times\nonumber\\
\times~{~} {\left(\frac{E_{iso}}{10^{53}
ergs}\right)}^{-1/8}{\left(\frac{\eta_\gamma}{0.2}\right)}^{1/8}{\left(\frac{n}{0.1cm^{-3}}\right)}^{1/8}~{~{~{~}}}~.~
\end{eqnarray}
where $\Theta_{j}$ is measured in radians. It was argued that the
fireball converts the energy in the ejecta into gamma rays
efficiently~\cite{Beloborodov00} (${\eta_\gamma}\approx${0.2}), and
that the mean circumburst density is {n}$\approx${0.1~cm$^{-3}$}
~\cite{Frail00}. Evaluating equation ~\ref{Theta} for the parameters
of GRB030329 (t$_{j}\approx$~10~hours, $z$=0.1685, and
E$_{iso}$=2$\times$10$^{52}$~erg) gives $\Theta_{j}\approx$0.07~
rad.

Therefore the beaming factor that relates the actual energy released
in gamma rays (E$_\gamma$) to the isotropic equivalent energy is
{${\Theta_{j}^{2}}$/2}$\approx$1/400, so that
E$_\gamma\approx$5$\times$10$^{49}$~erg. Comparing E$_{iso}$ and
E$_\gamma$ with the histograms in Fig. 2 of Frail \emph{et
al.}~\cite{Frail01} , GRB030329 resides at the lower end of the
energy distributions. The calculated isotropic energy from
GRB980425, the GRB associated with SN1998bw, is also low
(${\simeq}$10$^{48}$~erg).

\section{OVERVIEW OF THE LIGO DETECTORS}\label{sec:LIGO_Detectors}

The three LIGO detectors  are orthogonal arm Michelson laser
interferometers, aiming to detect gravitational waves by
interferometrically monitoring the relative (differential)
separation of mirrors, which play the role of test masses. The LIGO
Hanford Observatory (LHO) operates two identically oriented
interferometric detectors, which share a common vacuum envelope: one
having 4 km long arms (H1), and one having 2 km long arms (H2). The
LIGO Livingston Observatory operates a single 4 km long detector
(L1). The two sites are separated by $\simeq$3000~km, representing a
maximum arrival time difference of ${\simeq}\pm$10~ms.

A complete description of the LIGO interferometers as they were
configured during LIGO's first Science Run (S1) can be found in Ref
~\cite{Stan03}.

\subsection{Detector calibration and configuration}\label{sec:Configurations}

To calibrate the error signal, the response to a known
differential arm strain is measured, and the frequency-dependent
effect of the feedback loop gain is measured and compensated for.
During detector operation, changes in calibration are tracked by
injecting continuous, fixed-amplitude sinusoidal excitations into
the end test mass control systems, and monitoring the amplitude of
these signals at the measurement (error) point.  Calibration
uncertainties at the Hanford detectors were estimated to be
$<11\%$.

Significant improvements were made to the LIGO detectors following
the S1 run, held in early fall of 2002:

\begin{enumerate}
    \item {The analog suspension controllers on the H2 and L1
interferometers were replaced with digital suspension controllers
of the type installed on H1 during S1, resulting in lower
electronics noise.}
    \item {The noise from the optical lever servo that damps the angular
excitations of the interferometer optics was reduced.}
    \item {The wavefront sensing system for the H1 interferometer was
used to control 8 of 10 alignment degrees of freedom for the main
interferometer.  As a result, it maintained a much more uniform
operating point over the run.}
    \item {The high frequency sensitivity was improved by operating the
interferometers with higher effective power, about 1.5 W.}
\end{enumerate}

\begin{figure}[!t]
\includegraphics[angle=0,width=94mm]{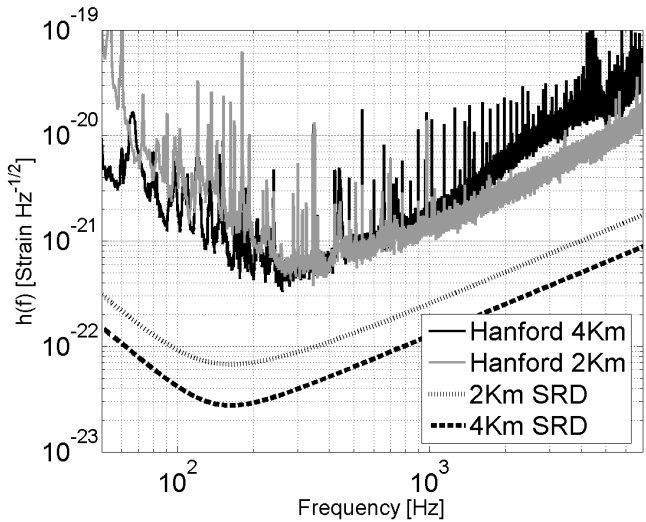}
\caption{Typical LIGO Hanford sensitivity curves during the S2 Run
[strain~Hz$^{-1/2}$] (black and grey lines). The LIGO design
sensitivity goals (SRD) are also indicated (dotted and dashed
lines).} \label{fig:LIGOEFF}
\end{figure}

These changes led to a significant improvement in detector
sensitivity. Figure~\ref{fig:LIGOEFF} shows typical spectra
achieved by the LIGO interferometers during the S2 run.
The differences among the three LIGO spectra reflect differences in
the operating parameters and hardware implementations of the three
instruments which are in various stages of reaching the final design
configuration.

\subsection{The second science run}\label{sec:S2Run}

The data analyzed in this paper were taken during LIGO's second
Science Run (S2), which spanned approximately 60 days from
February 14 to April 14, 2003. During this time, operators and
scientific monitors attempted to maintain continuous low noise
operation. The duty cycle for the interferometers, defined as the
fraction of the total run time when the interferometer was locked
and in its low noise configuration, was approximately 74\% for H1
and 58\% for H2.  The longest continuous locked stretch for any
interferometer during S2 was 66 hours for H1.

At the time of the GRB030329 both Hanford interferometers were
locked and taking science mode data. For this analysis we relied on
the single, ${\simeq}4.5$ hours long coincident lock stretch, which
started ${\simeq}3.5$ hours before the trigger time. With the
exception of the signal region, we utilized $\simeq$98\% of the data
within this lock stretch as the \emph{background} region (defined in
section V). 60~seconds of data before and after the signal region
were not included in the background region. Data from the beginning
and from the end of the lock stretch were not included in the
background region to avoid using possibly non-stationary data, which
might be associated with these regions.

As described below, the false alarm rate estimate, based on
background data, must be applicable to the data within the signal
region. We made a conservative choice and avoided using background
data outside of the lock stretch containing the GRB trigger time.
This is important when considering the present non-stationary
behavior of the interferometric detectors.

\section{Analysis}\label{sec:Analysis}

The goal of the analysis is either to identify significant events
in the signal region or, in the absence of significant events, to
set a limit on the strength of the associated gravitational wave
signal. Simulations and background data were used to determine the
detection efficiency for various ad-hoc and model-based waveforms
(Section~\ref{sec:Efficiencies}) and the false alarm rate of the
detection algorithm respectively.

The analysis takes advantage of the information provided by the
astrophysical trigger. The trigger time determined when to perform
the analysis. As discussed below, the time window to be analyzed
around the trigger time was chosen to accommodate most current
theoretical predictions and timing uncertainties. The source
direction was needed to calculate the attenuation due to the LIGO
detector's antenna pattern for the astrophysical interpretation.

The two co-located and co-aligned Hanford detectors had very similar
frequency-dependent response functions at the time of the trigger.
Consequently, the detected arrival time and recorded waveforms of a
gravitational wave signal should be essentially the same in both
detectors. It is natural then to consider cross-correlation of the
two data streams as the basis of a search algorithm. This conclusion
can also be reached via a more formal argument based on the maximum
log-Likelihood ratio test~\cite{Anderson01,Mohanty04}.

\begin{figure}[!t]
\includegraphics[angle=0,width=84mm]{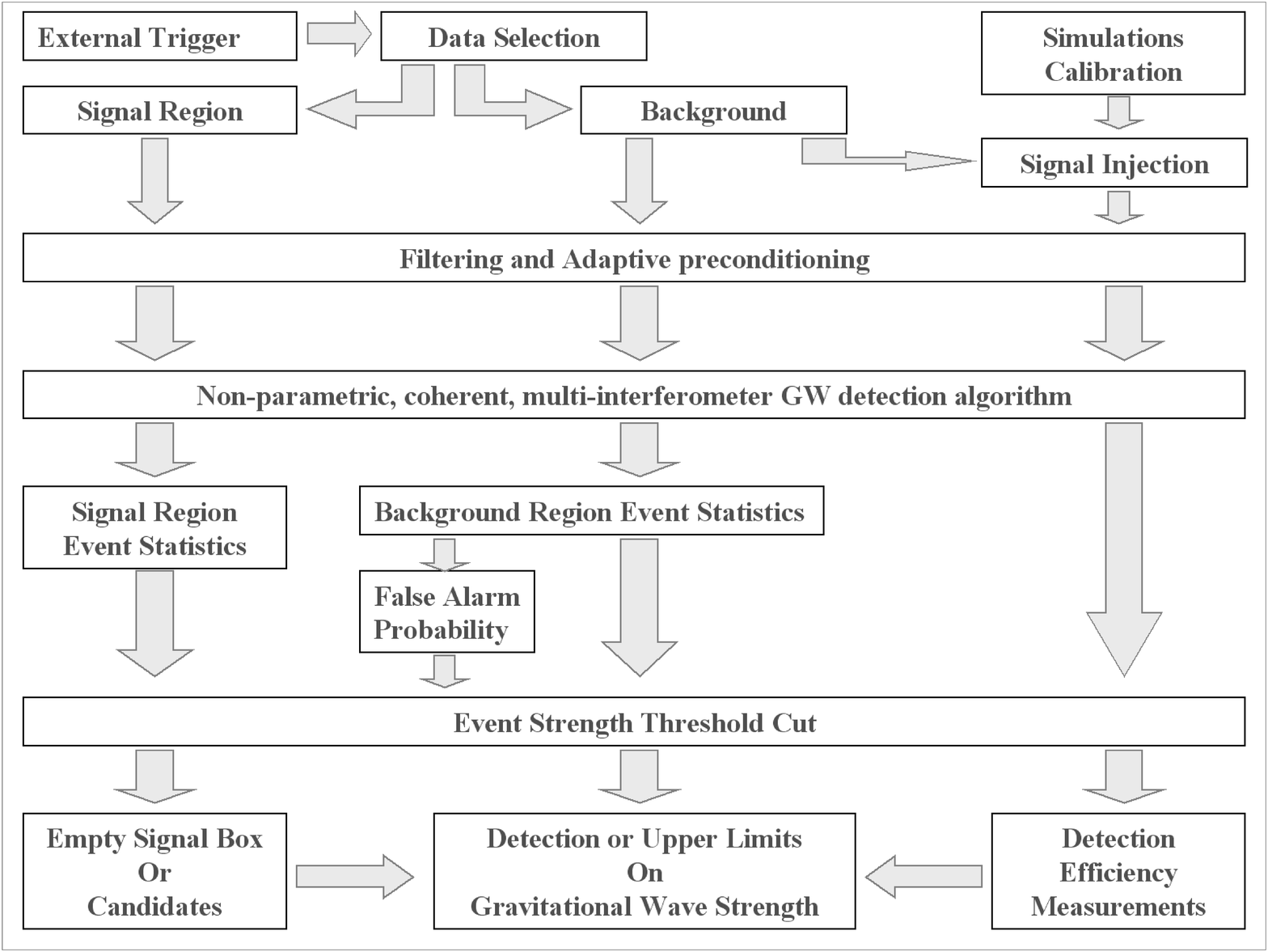}
\caption{The schematic of the analysis pipeline } \label{fig:PIPE}
\end{figure}

The schematic of the full analysis pipeline is shown in
Figure~\ref{fig:PIPE}. The underlying analysis algorithm is
described in detail in Ref.~\cite{Mohanty04}.
The background data, the signal region data and the simulations are
all processed identically. The background region consists of the
data where we do not expect to have a gravitational wave signal
associated with the GRB. We scan the background to determine the
false alarm distribution and to set a threshold on the event
strength that will yield an acceptable false alarm rate. This
threshold is used when scanning the signal region and simulations.
In order to estimate our sensitivity to gravitational waves,
simulated signals of varying strength are added to the detector data
streams.

The signal region around the GRB trigger is scanned to identify
outstanding signals. If events were detected above threshold, in
this region, their properties would be tested against those
expected from gravitational waves. If no events were found above
threshold, we would use the estimated sensitivity to set an upper
limit on the gravitational wave strain at the detector.

The output from each interferometer is divided into 330~sec long
segments with a 15~second overlap between consecutive segments (both
ends), providing a tiling of the data with 300~second long segments.
In order to avoid edge effects, the 180~sec long signal region lies
in the middle of one such 300~sec long segment. This tiling method
also allows for adaptive data conditioning and places the
conditioning filter (see Sec.~\ref{sec:DataConditioning} below)
transients well outside of the 300 second long segment containing
the signal region.

\subsection{Choice of signal region}\label{sec:Signal_region}

Current models suggest~\cite{Meszaros03} that the gravitational wave
signature should appear close to the GRB trigger time. We
conservatively chose the duration and position of the signal region
to over-cover most predictions and to allow for the expected
uncertainties associated with the GRB trigger timing. A 180 second
long window (see Figure~\ref{fig:HETE}), starting 120 seconds before
the GRB trigger time is sufficient; roughly ten times wider than the
GRB light curve features, and wide enough to include most
astrophysical predictions. Most models favor an ordering where the
arrival of the gravitational wave precedes the GRB
trigger~\cite{Meszaros03}, but in a few other cases the
gravitational wave arrival is predicted to be
contemporaneous~\cite{Araya-Gochez03,Putten04} to the arrival and
duration of the gamma rays (i.e after the GRB trigger). The 60
second region after the GRB trigger time, is sufficient to cover
these predictions and also contains allowance for up to 30 seconds
uncertainty on trigger timing, which is a reasonable choice in the
context of the HETE light curve. Figure~(\ref{fig:HETE}) shows a
signal rise time of order $10~s$, precursor signals separated from
the main peak, and significant structure within the main signal
itself. Effects due to the beaming dynamics of the GRB and the
instrumental definition of the trigger time can also be significant
contributors to the timing uncertainty.

\subsection{Search algorithm}\label{sec:SearchAlgorithm}

\subsubsection{Data Conditioning}\label{sec:DataConditioning}

The data-conditioning step was designed to remove instrumental
artifacts from the data streams. We used an identical data
conditioning procedure when processing the background, the signal
region and the simulations.

The raw data streams have narrowband lines, associated with the
power line harmonics at multiples of 60Hz, the violin modes of the
mirror suspension wires and other narrow band noise sources. The
presence of lines has a detrimental effect on our sensitivity
because lines can produce spurious correlations between detectors.
In addition, the broadband noise shows significant variations over
timescales of hours and smaller variations over timescales of
minutes and seconds due to alignment drift and fluctuations. The
background data must portray a representative sample of the
detector behavior around the time of the trigger. Broadband
non-stationarity can limit the duration of this useful background
data and hence the reliability of our estimated false alarm rate.

Our cross correlation based algorithm performs best on white
spectra without line features. We use notch filters to remove the
well-known lines, such as power line and violin mode harmonics
from both data streams. Strong lines of unknown origin with
stationary mean frequency are also removed at this point. We also
apply a small correction to mitigate the difference between the
phase and amplitude response of the two Hanford detectors.

We bandpass filter and decimate the data to a sampling rate of
4096~Hz to restrict the frequency content to the ${\simeq}80$~Hz to
${\simeq}2048$~Hz region, which was the most sensitive band for both
LIGO Hanford detectors during the S2 run.

In order to properly remove weaker stationary lines and the small
residuals of notched strong lines, correct for small slow changes
in the spectral sensitivity and whiten the spectrum of the data,
we use adaptive line removal and whitening. As all strong lines
are removed before the adaptive whitening, we avoid potential
problems due to non-stationary lines and enhance the efficiency of
the follow up adaptive filtering stage. The conditioned data has a
consistent white spectrum without major lines and sufficient
stationarity, from segment to segment, throughout the background
and signal regions.

The end result of the pre-processing is a data segment with a flat
power spectral density (white noise), between $\simeq$80~Hz and
$\simeq$2048~Hz. The data conditioning was applied consistently
after the signal injections. This ensures that any change in
detection efficiency due to the pre-processing is properly taken
into account.

\subsubsection{Gravitational Wave Search Algorithm}\label{sec:Algorithm}

The test statistics for a pair of data streams are constructed as
follows. We take pairs of short segments, one from each stretch,
and compute their cross-correlation function. The actual form of
the cross-correlation used ($\widetilde{C}^{m,n}_{k,p,j}$) is
identical to the common Euclidean inner product:
\begin{equation}
\widetilde{C}^{m,n}_{k,p,j}= \sum_{i=-j}^{j} H_m[k+i]
H_n[k+p+i]~,~
\end{equation}
where the pre-conditioned time series from detector ``x" is ${\bf
H}_x=\{H_x[0],H_x[1],\ldots\}$ and i,k,p and j are all integers
indexing the data time series, with each datum being (1/4096)~s
long. As we now only consider the two Hanford detectors ``m" and
``n" can only assume values of 1 ($H_1$) or 2 ($H_2$). There are
therefore three free parameters to scan when searching for
coherent segments of data between a pair of interferometers (m,n):
1. the center time of the segment from the first detector (k); 2.
the relative time lag between the segments from the two detectors
(p); and 3. the common duration of segments (2j+1) called the
integration length.

The optimum integration length to use for computing the
cross-correlation depends on the duration of the signal and its
signal-to-noise ratio, neither of which is {\em a priori} known.
Therefore the cross-correlation should be computed from segment
pairs with start times and lengths varying over values, which
should, respectively, cover the expected arrival times (signal
region) and consider durations of the gravitational wave burst
signals~\cite{Muller2004,Fryer2004,Fryer2002,Zwerger1997,Dimmelmeier2002,Burrows1996}
($\sim$O(1-128ms)).

Hence we apply a search algorithm~\cite{Mohanty04} that processes
the data in the following way.

(1) A three dimensional quantity (${\cal C}_{k,j}[p]$) is
constructed:
\begin{equation}\label{CCStat}
{\cal C}_{k,j}[p] =
\left[\left({\widetilde{C}^{1,2}_{k,p,j}}\right)^2 +
\left({\widetilde{C}^{2,1}_{k,-p,j}}\right)^2\right]^{1/2}\;,
\end{equation}
scanning the range of segment center times (k), integration
lengths (2j+1) and relative time shifts ($p =
0,\pm1,\pm2,\ldots$). A coherent and coincident signal is expected
to leave its localized signature within this three dimensional
quantity.

We use a fine rectangular grid in relative time shift (p) and
integration length (2j+1) space. The spacing between grid points is
$\simeq$1~ms for the segment center time (k) and (1/4096)~s for the
relative time shift. The spacing of the integration lengths is
approximately logarithmic. Each consecutive integration length is
${\simeq}$50\% longer than the previous one, covering integration
lengths from ${\simeq}$1~ms to ${\simeq}$128~ms.

Introducing small, non-physical relative time shifts (much larger
than the expected signal duration) between the two data streams
before computing the cross-correlation matrix suppresses the average
contribution from a GW signal. This property can be used to estimate
the local noise properties, thereby mitigating the effects of
non-stationarity in the interferometer outputs. Accordingly, ${\cal
C}_{k,j}[p]$ contains the autocorrelation of the coherent signal for
relative time shifts at and near p~$=$~0 (called ``\emph{core}"),
while far away, in the ``side \emph{lobes}", the contribution from
the signal autocorrelation is absent, sampling only the random
contributions to the cross-correlation arising from the noise. The
optimal choice of the core size depends on the expected signal
duration (integration length), the underlying detector noise and it
cannot be smaller than the relative phase uncertainty of the
datastreams. The core region can reach as far as 5~ms, as it
increases with increasing integration length. The size of each side
lobe is twice the size of the core region and the median time shift
associated with the side lobes can be as large as 325~ms as it is
also increasing with increasing integration length. We use the side
lobes of ${\cal C}_{k,j}[p]$ to estimate the mean
($\widehat{\mu}_{k,j}$) and variance ($\widehat{\sigma}_{k,j}$) of
the local noise distribution, which is also useful in countering the
effects of non-stationarity.

\begin{figure}[!t]
\includegraphics[angle=0,width=87mm]{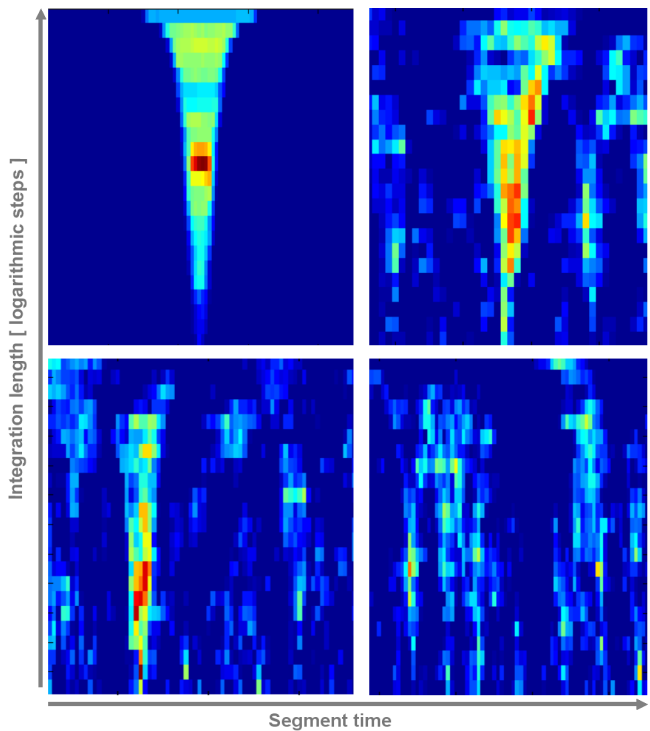}
\caption{Examples of corrgram images. The horizontal axes are time
(linearly scaled) and the vertical axes are integration length
(logarithmically scaled). The color axis, an indicator of the excess
correlation, is independently auto-scaled for each quadrant for
better visibility, therefore the meaning of colors differ from
quadrant to quadrant. The time ticks also change from quadrant to
quadrant for better visibility. The rainbow type color scale goes
from blue to red, dark red marking the most significant points
within a quadrant. The upper two quadrants show the corrgram image
of injected Sine-Gaussians (250~Hz,~Q~$=$~8.9). The bottom quadrants
are examples of noise. The maximum of the intensity scale is
significantly higher for both quadrants with injections, when
compared to the noise examples. The top left injection is strong
enough to be significantly above the preset detection threshold,
while the top right injection is weak enough to fall significantly
below the detection threshold.} \label{fig:Corrgram}
\end{figure}

(2) The three dimensional quantity is reduced to a two dimensional
image (see Fig.~\ref{fig:Corrgram}), called a {\em corrgram}, as
follows. The values of ${\cal C}_{k,j}[p]$ in the core region are
standardized by subtracting $\widehat{\mu}_{k,j}$ and then dividing
by $\widehat{\sigma}_{k,j}$. {\em Positive} standardized values in
the core region are summed over $p$ to determine the value of the
corrgram pixel. Each pixel is a measure of the excess
cross-correlation in the core region when compared to the expected
distribution characterized by the side lobes for the given (k,j)
combination.

(3) A list of events is found by recursively identifying and
characterizing significant regions (called ``clusters") in the
corrgram image. Each event is described by its arrival time, its
optimal integration length and its strength (ES). The event's
arrival time and its optimal integration length correspond to the
most significant pixel of the cluster. The event strength is
determined by averaging the five most significant pixels of the
cluster, as this is helpful in discriminating against random
fluctuations of the background noise.

The strength of each event is then compared to a preset detection
threshold corresponding to the desired false alarm rate. This
detection threshold is determined via extensive scans of the
background region.

\section{Results}\label{sec:Results}

\subsection{False alarm rate measurements}\label{sec:FAR}

In order to assess the significance of the cross-correlated power
of an event, we determined the false alarm rate versus event
strength distribution. We used the full background data stretch
for this measurement.

\begin{figure}[!t]
\includegraphics[angle=0,width=92mm]{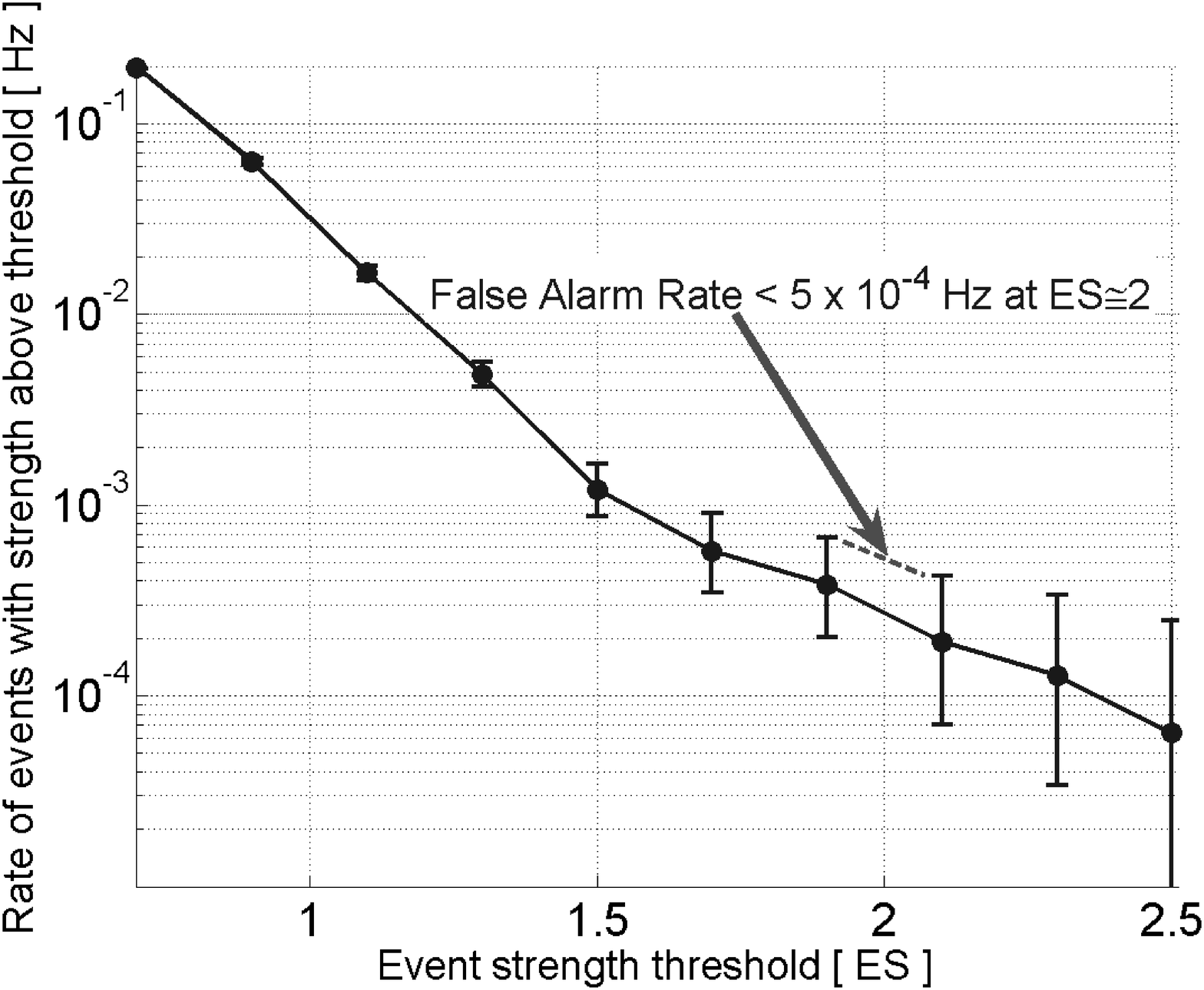}
\caption{False alarm rate as a function of the event strength
threshold as determined from background data. The error bars reflect
90\% CL Poisson errors, based on the the number of events within the
given bin. The pointer indicates the event strength threshold used
for the analysis, which corresponds to an interpolated false alarm
rate of less than $5\times10^{-4}$~Hz. Note that the signal region
data is not included in this calculation. The position of symbols
correspond to the center of the bins.} \label{fig:FAR}
\end{figure}

Figure~\ref{fig:FAR} shows the event rate as a function of the
event strength threshold for the background region. The error bars
reflect 90\% CL Poisson errors, based on the the number of events
within the given bin. We used this distribution to fix the event
strength threshold used in the subsequent analysis.

We chose an event strength threshold with an associated false alarm
rate of less than ${\simeq }$5$\times$10$^{-4}$~Hz, equivalent to
less than ${\simeq}$9\% chance for a false alarm within the 180
second long signal region.

\subsection{Efficiency determination}\label{sec:Efficiencies}

The detection sensitivity of the analysis was determined by
simultaneously adding simulated signals of various amplitudes and
waveforms to both data streams in the background region and
evaluating the efficiency of their detection as a function of the
injected amplitude and waveform type.

The waveforms we considered include Sine-Gaussians to emulate
short narrow-band bursts, Gaussians to emulate short broad-band
signals, and Dimmelmeier-Font-M\"{u}ller numerical
waveforms~\cite{Dimmelmeier2002}, as examples of astrophysically
motivated signals.

Calibration of the waveforms from strain to ADC counts was
performed in the frequency domain, and was done separately for
each interferometer. Calibration procedures of the LIGO detectors
are described in Ref.~\cite{Stan03}. The transformed signals, now
in units of counts of raw interferometer noise, were then simply
added to the raw data stream.

The amplitudes and the times of the injections were randomly
varied. In this way we ensured that each amplitude region sampled
the full, representative range of noise variations and that we had
no systematic effects, for example, due to a regular spacing in
time.

To a reasonable approximation the sensitivity of our analysis
pipeline can be expressed in terms of the frequency content, the
duration and the strength of the gravitational wave signal.
Therefore, it is sufficient to estimate the sensitivity of our
search for a representative set of broad and narrow band waveforms,
which span the range of frequencies, bandwidth, and duration we wish
to search.

We characterize the strength of an arbitrary waveform by its
root-sum-square amplitude ($h_{RSS}$), which is defined
as~\cite{Burstpaper}:
\begin{equation}\label{hRSS}
h_{RSS} = \sqrt{\int_{-\infty}^{\infty} |~h(t)~|^2~ \,dt }~.~
\end{equation}
The above definition of $h_{RSS}$ includes all frequencies, while
the gravitational wave detectors and search algorithms are only
sensitive in a restricted frequency band. In principle, one can
analogously define a ``band-limited" $h_{RSS}$, in which only the
sensitive frequency band of the analysis is taken into account.
Within this paper we choose to adopt the Eq.~\ref{hRSS} definition
of h$_{RSS}$ for historical reasons.

\begin{figure}[!t]
\includegraphics[angle=0,width=84mm]{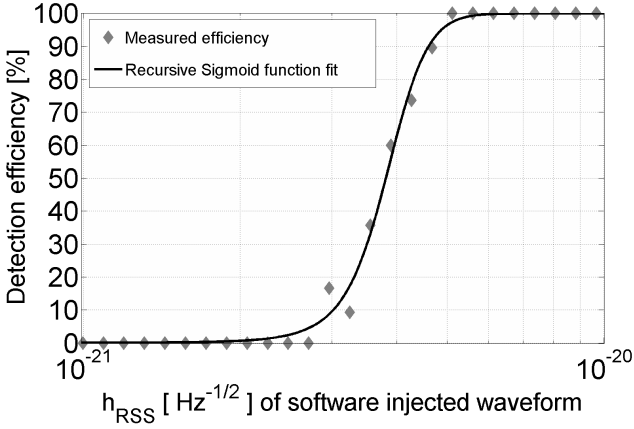}
\caption{Efficiency of the detection algorithm for a sample waveform
as a function of signal strength ($h_{RSS}$); in this case a
Sine-Gaussian of $f_0=250$~Hz and Q~$=$~8.9. To extract this curve
numerous simulated waveforms were embedded in a representative
fraction of the background data at random times with randomly
varying signal strength. The plot shows the fraction of signals
detected as the function of amplitude and a sigmoid function fit.
The reconstructed signal onset times were required to fall within
$\pm$60~ms of the true onsets, which also explains why the low
$h_{RSS}$ end of the curve falls near zero. This is a typical plot
and in general, the agreement between the measured values, and the
fit is better than ${\simeq}5\%$. We relied on the fit to extract
our upper limits for an optimally oriented and polarized source.
Section~\ref{sec:Interpretation} below describes the corrections due
to non-optimal source direction and polarization.}
\label{fig:EFF_250_8.9}
\end{figure}

\begin{figure}[!t]
\includegraphics[angle=0,width=84mm]{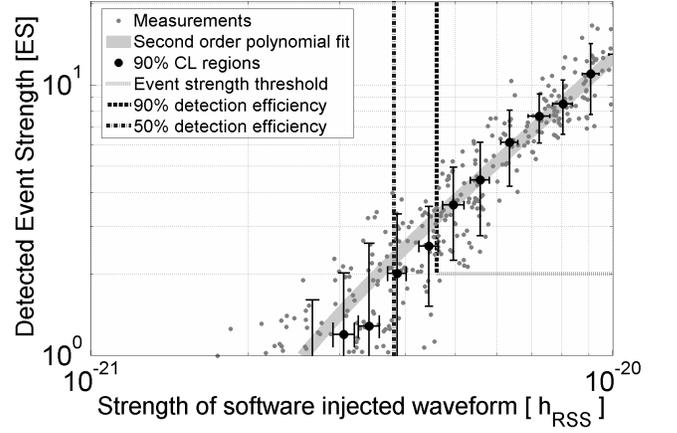}
\caption{Detected event strength versus $h_{RSS}$ of the injected
Sine-Gaussian waveform with $f_0=250$~Hz and $Q~\approx~8.9$. The
dots indicate the scatter of the distribution of raw measurements.
The gray band shows the quadratic polynomial fit, which allows us to
convert the strength of an observed event into the equivalent
$h_{RSS}$ value and determine the associated $90\%$ CL error bars.
The markers with error bars represent the $90\%$ CL regions for
subsets of the data. For each marker, $90\%$ of the measurements
used were within the horizontal error bars and $90\%$ of the
detected event strengths values fell within the vertical error bars.
The vertical dash-dot line represents the $50\%$ detection
efficiency associated with the waveform type and the chosen
detection threshold (horizontal dotted line). As expected, the
crossing of the threshold and the $50\%$ efficiency lines agree well
with the fit and the center of the corresponding marker. The
vertical dashed line represents the boundary of the region where we
have better than $90\%$ detection efficiency. The ``corner" defined
by the event strength threshold and the $90\%$ detection efficiency
boundary (dashed lines) agrees well with the curve outlined by the
lower end of the vertical error bars of the markers. All events in
the upper right corner of the plot (above and beyond the dashed
lines) are detectable with high confidence. This plot is typical for
different waveforms considered in the analysis.}
\label{fig:CAL_250_8.9}
\end{figure}

The extracted sensitivities (see the examples in
Figures~\ref{fig:EFF_250_8.9} and ~\ref{fig:CAL_250_8.9}) can be
used to generalize our measurements and estimate the pipeline's
sensitivity for other similar band-limited waveforms.

\begin{figure}[!t]
\includegraphics[angle=0,width=94mm]{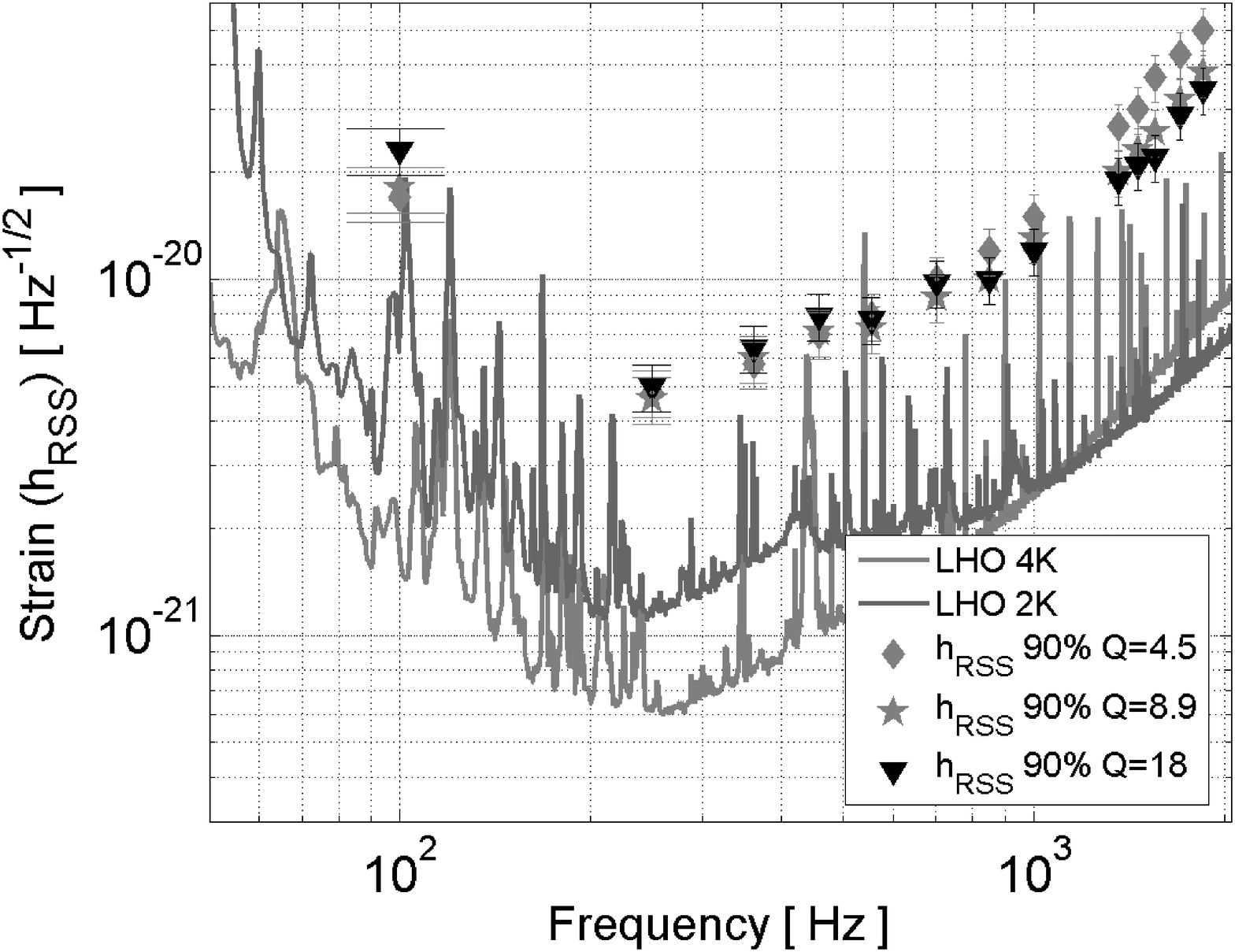}
\caption{Sensitivity of the detection algorithm for detecting
Sine-Gaussian waveforms versus characteristic frequency. The plot
shows the strength necessary for $90\%$ detection efficiency. The
grey spectra illustrate the sensitivity of the 2K and 4K Hanford
detectors during the time surrounding the GRB030329 trigger. The
error bars reflect a total $15\%$ error.} \label{fig:SENS}
\end{figure}

To assess the sensitivity for relatively narrow band waveforms, we
used Sine-Gaussian injections of the form:

\begin{equation}\label{SineGauss}
h(t) = h_\circ \sin(\omega_\circ\,t)e^{-t^2/2\sigma^2} \;,
\end{equation}
with a central angular frequency of~~$\omega_\circ = 2\pi
f_\circ$, and $Q=\omega_\circ \sigma = 2\pi f_\circ \sigma $. The
relation between $h_\circ$ and $h_{RSS}$ is given as:

\begin{eqnarray}\label{SGhRSS}
h_{RSS}^{SG} = h_\circ~\sqrt{\frac{\sqrt{\pi}~\sigma}{2} \left(
1-e^{-Q^2}\right)}\, ~{\approx}~\nonumber\\
~{\approx}~ h_\circ~\sqrt{\frac{Q}{4\sqrt{\pi}~f_\circ}}~ \times
~\left\{
\begin{array}{cc}
  1 & Q\gg1 \\
  0.8 & Q\simeq1 \\
\end{array}
\right\}\;.
\end{eqnarray}
\\

The injected signals covered the frequency range between 100 and
1850 Hz with 13 values of $f_{\circ}$. To test the dependence of the
sensitivity on signal duration, we used three values of Q (4.5, 8.9
and 18) for each frequency (see Table~\ref{t:NBET}). Near the most
sensitive frequency region, around $\simeq$250~Hz, our gravitational
wave strain sensitivity for optimally polarized bursts was better
than h$_{RSS}{\simeq}$5$\times$10$^{-21}$~Hz$^{-1/2}$.
Figure~\ref{fig:SENS} shows the sensitivity for these narrow band
waveforms. The symbols mark the simulated event strength ($h_{RSS}$)
necessary to achieve $90\%$ detection efficiency for each waveform.
We quote the gravitational wave signal strength associated with the
$90\%$ detection efficiency, as this can be related to the upper
limits on the gravitational wave strength associated with the
source. Figure~\ref{fig:SENS} also illustrates the insensitivity of
the detection efficiency to the Q of the Sine-Gaussian waveforms
with the same central frequency, as these reach their $90\%$
efficiency levels at similar gravitational wave strengths, even
though their Q differ by a factor of ${\simeq}4$; for a given
h$_{RSS}$, a longer signal (higher Q) would of course, have a
smaller h$_{PEAK}$. This strength is frequency dependent, naturally
following the frequency dependence of the detector sensitivities,
which are also indicated in Figure~\ref{fig:SENS}.

Table~\ref{t:BBET} shows a similar set of efficiencies estimated
using broad-band simulated signals. We used two types of broadband
waveforms, Sine-Gaussians with unity quality factor and Gaussians.
Both are short bursts, however, the Gaussians are even functions
while the Sine-Gaussians are odd, leading to different peak
amplitudes with the same $h_{RSS}$ value. Gaussians were
parametrized as:

\begin{equation}\label{Gaussian}
h(t) = h_\circ e^{-t^2/2\sigma^2} \;.
\end{equation}

The relationship between $h_\circ$ and $h_{RSS}$ for a Gaussian
is:

\begin{equation}\label{GaussianhRSS}
h_{RSS}^{GA} = h_\circ~\sqrt{ \sqrt{\pi}\>\sigma}\;.
\end{equation}

The estimated sensitivities indicate that the 90$\%$ detection
efficiency limits for short bursts are similar to those obtained for
the narrow band waveforms when one takes into account that only part
of the power of the broad-band waveforms is confined to the analysis
frequency band. Longer Gaussian bursts are more difficult to detect,
as their spectrum has a significant low frequency component, outside
the sensitive band of our analysis.

We have also estimated our efficiency for a set of astrophysically
motivated burst waveforms~\cite{Dimmelmeier2002} (see Table
~\ref{t:DFMT}). These simulated waveforms are not expected to be
necessarily associated with GRBs, rather these results are presented
here to further illustrate the waveform independence of the
analysis.

\begin{table} [htb]
\begin{center}
\caption[]{\label{t:NBET} $h_{RSS}$ [$Hz^{-1/2}$] for 90\% detection
efficiency for Sine-Gaussians (SG) waveforms at various frequencies
($f_{\circ}$) and Q (see eq.~\ref{SineGauss}). The quoted values are
the results of simulations and are subject to a total of
${\simeq}~15\%$ statistical and systematic errors, which are taken
into account when quoting the ${UL}^{90\%CL}_{h_{RSS}}$ values. Note
that at the low and at the high frequency end, the low Q waveforms
have significant power outside of the analysis frequency band.}
\small
\begin{tabular}{lccccc}
\hline\hline
Waveform & $\sigma$ [ms] & Q & $f_{\circ} [Hz]$ & $h_{RSS}^{90\%} [Hz^{-1/2}]$ & ${UL}^{90\%CL}_{h_{RSS}} [Hz^{-1/2}]$ \\
\hline
SG & 7.2 & 4.5 & 100 & $17\times 10^{-21}$ & $20\times 10^{-21}$ \\
SG & 2.9 & 4.5 & 250 & $4.8\times 10^{-21}$ & $5.6\times 10^{-21}$ \\
SG &  2 & 4.5 & 361 & $5.8\times 10^{-21}$ & $6.7\times 10^{-21}$ \\
SG & 1.6 & 4.5 & 458 & $ 7.0\times 10^{-21}$ & $ 8.0\times 10^{-21}$ \\
SG & 1.3 & 4.5 & 554 & $7.9\times 10^{-21}$ & $9.1\times 10^{-21}$ \\
SG &  1 & 4.5 & 702 & $10\times 10^{-21}$ & $11\times 10^{-21}$ \\
SG & 0.84 & 4.5 & 850 & $12\times 10^{-21}$ & $14\times 10^{-21}$ \\
SG & 0.72 & 4.5 & 1000 & $15\times 10^{-21}$ & $17\times 10^{-21}$ \\
SG & 0.53 & 4.5 & 1361 & $27\times 10^{-21}$ & $31\times 10^{-21}$ \\
SG & 0.49 & 4.5 & 1458 & $30\times 10^{-21}$ & $34\times 10^{-21}$ \\
SG & 0.46 & 4.5 & 1554 & $37\times 10^{-21}$ & $43\times 10^{-21}$ \\
SG & 0.42 & 4.5 & 1702 & $43\times 10^{-21}$ & $50\times 10^{-21}$ \\
SG & 0.39 & 4.5 & 1850 & $50\times 10^{-21}$ & $58\times 10^{-21}$ \\
\hline
SG & 14 & 8.9 & 100 & $18\times 10^{-21}$ & $21\times 10^{-21}$ \\
SG & 5.7 & 8.9 & 250 & $4.6\times 10^{-21}$ & $5.3\times 10^{-21}$ \\
SG & 3.9 & 8.9 & 361 & $ 6.0\times 10^{-21}$ & $6.9\times 10^{-21}$ \\
SG & 3.1 & 8.9 & 458 & $7.1\times 10^{-21}$ & $8.1\times 10^{-21}$ \\
SG & 2.6 & 8.9 & 554 & $7.3\times 10^{-21}$ & $8.4\times 10^{-21}$ \\
SG &  2 & 8.9 & 702 & $8.9\times 10^{-21}$ & $10\times 10^{-21}$ \\
SG & 1.7 & 8.9 & 850 & $10\times 10^{-21}$ & $12\times 10^{-21}$ \\
SG & 1.4 & 8.9 & 1000 & $13\times 10^{-21}$ & $15\times 10^{-21}$ \\
SG &  1 & 8.9 & 1361 & $20\times 10^{-21}$ & $23\times 10^{-21}$ \\
SG & 0.97 & 8.9 & 1458 & $23\times 10^{-21}$ & $27\times 10^{-21}$ \\
SG & 0.91 & 8.9 & 1554 & $26\times 10^{-21}$ & $30\times 10^{-21}$ \\
SG & 0.83 & 8.9 & 1702 & $32\times 10^{-21}$ & $37\times 10^{-21}$ \\
SG & 0.77 & 8.9 & 1850 & $38\times 10^{-21}$ & $44\times 10^{-21}$ \\
\hline
SG & 29 & 18 & 100 & $23\times 10^{-21}$ & $26\times 10^{-21}$ \\
SG & 11 & 18 & 250 & $ 5.0\times 10^{-21}$ & $5.7\times 10^{-21}$ \\
SG & 7.9 & 18 & 361 & $6.4\times 10^{-21}$ & $7.4\times 10^{-21}$ \\
SG & 6.3 & 18 & 458 & $7.9\times 10^{-21}$ & $9.1\times 10^{-21}$ \\
SG & 5.2 & 18 & 554 & $7.7\times 10^{-21}$ & $8.9\times 10^{-21}$ \\
SG & 4.1 & 18 & 702 & $9.8\times 10^{-21}$ & $11\times 10^{-21}$ \\
SG & 3.4 & 18 & 850 & $10\times 10^{-21}$ & $12\times 10^{-21}$ \\
SG & 2.9 & 18 & 1000 & $12\times 10^{-21}$ & $14\times 10^{-21}$ \\
SG & 2.1 & 18 & 1361 & $19\times 10^{-21}$ & $21\times 10^{-21}$ \\
SG &  2 & 18 & 1458 & $21\times 10^{-21}$ & $24\times 10^{-21}$ \\
SG & 1.8 & 18 & 1554 & $22\times 10^{-21}$ & $25\times 10^{-21}$ \\
SG & 1.7 & 18 & 1702 & $29\times 10^{-21}$ & $33\times 10^{-21}$ \\
SG & 1.5 & 18 & 1850 & $34\times 10^{-21}$ & $39\times 10^{-21}$ \\
\hline\hline
\end{tabular}
\end{center}
\end{table}

\begin{table} [htb]
\begin{center}
\caption[]{\label{t:BBET} As in Table~\ref{t:NBET}, $h_{RSS}$
[$Hz^{-1/2}$] for 90\% detection efficiency for Gaussian (GA)
waveforms of various durations ($\sigma$) (see eq.~\ref{Gaussian})
and for Sine-Gaussians (SG) waveforms at various frequencies
($f_{\circ}$) and $Q=1$ (see eq.~\ref{SineGauss}). Note that these
broadband waveforms have significant power outside of the analysis
frequency band.} \small
\begin{tabular}{lccccc}
\hline\hline
Waveform & $\sigma$ [ms] & Q & $f_{\circ} [Hz]$ & $h_{RSS}^{90\%} [Hz^{-1/2}]$ & ${UL}^{90\%CL}_{h_{RSS}} [Hz^{-1/2}]$ \\
\hline
SG & 1.6   &  1 & 100  & $10\times 10^{-21}$  & $12\times 10^{-21}$ \\
SG & 0.64  &  1 & 250  & $6.5\times 10^{-21}$ & $7.4\times 10^{-21}$ \\
SG & 0.44  &  1 & 361  & $8.4\times 10^{-21}$ & $9.7\times 10^{-21}$ \\
SG & 0.35  &  1 & 458  & $10\times 10^{-21}$  & $12\times 10^{-21}$ \\
SG & 0.29  &  1 & 554  & $13\times 10^{-21}$  & $14\times 10^{-21}$ \\
SG & 0.23  &  1 & 702  & $18\times 10^{-21}$  & $20\times 10^{-21}$ \\
SG & 0.19  &  1 & 850  & $23\times 10^{-21}$  & $26\times 10^{-21}$ \\
SG & 0.16  &  1 & 1000 & $26\times 10^{-21}$  & $30\times 10^{-21}$ \\
SG & 0.12  &  1 & 1361 & $39\times 10^{-21}$  & $45\times 10^{-21}$ \\
SG & 0.11  &  1 & 1458 & $44\times 10^{-21}$  & $51\times 10^{-21}$ \\
SG & 0.1   &  1 & 1554 & $46\times 10^{-21}$  & $52\times 10^{-21}$ \\
SG & 0.094 &  1 & 1702 & $55\times 10^{-21}$  & $63\times 10^{-21}$ \\
SG & 0.086 &  1 & 1850 & $61\times 10^{-21}$  & $70\times 10^{-21}$ \\
\hline
GA & 0.5  & & & $8.3\times 10^{-21}$ & $9.6\times 10^{-21}$ \\
GA & 0.75 & & & $9.6\times 10^{-21}$ & $1.1\times 10^{-20}$ \\
GA &  1   & & & $1.3\times 10^{-20}$ & $1.5\times 10^{-20}$ \\
GA &  2   & & & $3.3\times 10^{-20}$ & $3.8\times 10^{-20}$ \\
GA &  3   & & & $8.2\times 10^{-20}$ & $9.5\times 10^{-20}$ \\
GA &  4   & & & $1.9\times 10^{-19}$ & $2.2\times 10^{-19}$ \\
GA & 5.5  & & & $8.5\times 10^{-19}$ & $9.8\times 10^{-19}$ \\
GA &  8   & & & $1.3\times 10^{-17}$ & $1.5\times 10^{-17}$ \\
GA & 10   & & & $1.0\times 10^{-16}$ & $1.2\times 10^{-16}$ \\
 \hline\hline
\end{tabular}
\end{center}
\end{table}

\begin{table} [htb]
\begin{center}
\caption[]{\label{t:DFMT} As in Table~\ref{t:NBET}, $h_{RSS}$
[$Hz^{-1/2}$] for 90\% detection efficiency for astrophysically
motivated waveforms. These waveforms are described in detail in
Ref.~\cite{Dimmelmeier2002}. Note that most of these waveforms
have significant power outside of the analysis frequency band.}
\small
\begin{tabular}{lccccc}
\hline\hline
Simulation & Waveform & $h_{RSS}^{90\%} [Hz^{-1/2}]$ & ${UL}^{90\%CL}_{h_{RSS}} [Hz^{-1/2}]$ \\
\hline
DFM & A1B1G1 & $12\times 10^{-21}$ & $14\times 10^{-21}$ \\
DFM & A1B2G1 & $13\times 10^{-21}$ & $15\times 10^{-21}$ \\
DFM & A1B3G1 & $12\times 10^{-21}$ & $14\times 10^{-21}$ \\
DFM & A1B3G2 & $12\times 10^{-21}$ & $14\times 10^{-21}$ \\
DFM & A1B3G3 & $12\times 10^{-21}$ & $14\times 10^{-21}$ \\
DFM & A1B3G5 & $34\times 10^{-21}$ & $39\times 10^{-21}$ \\
DFM & A2B4G1 & $24\times 10^{-21}$ & $27\times 10^{-21}$ \\
DFM & A3B1G1 & $19\times 10^{-21}$ & $21\times 10^{-21}$ \\
DFM & A3B2G1 & $20\times 10^{-21}$ & $23\times 10^{-21}$ \\
DFM & A3B2G2 & $15\times 10^{-21}$ & $17\times 10^{-21}$ \\
DFM & A3B2G4 & $14\times 10^{-21}$ & $16\times 10^{-21}$ \\
DFM & A3B3G1 & $28\times 10^{-21}$ & $33\times 10^{-21}$ \\
DFM & A3B3G2 & $17\times 10^{-21}$ & $20\times 10^{-21}$ \\
DFM & A3B3G3 & $12\times 10^{-21}$ & $14\times 10^{-21}$ \\
DFM & A3B3G5 & $30\times 10^{-21}$ & $34\times 10^{-21}$ \\
DFM & A3B4G2 & $23\times 10^{-21}$ & $27\times 10^{-21}$ \\
DFM & A3B5G4 & $26\times 10^{-21}$ & $29\times 10^{-21}$ \\
DFM & A4B1G1 & $38\times 10^{-21}$ & $44\times 10^{-21}$ \\
DFM & A4B1G2 & $32\times 10^{-21}$ & $36\times 10^{-21}$ \\
DFM & A4B2G2 & $42\times 10^{-21}$ & $48\times 10^{-21}$ \\
DFM & A4B2G3 & $39\times 10^{-21}$ & $45\times 10^{-21}$ \\
DFM & A4B4G4 & $17\times 10^{-21}$ & $19\times 10^{-21}$ \\
DFM & A4B4G5 & $12\times 10^{-21}$ & $13\times 10^{-21}$ \\
DFM & A4B5G4 & $21\times 10^{-21}$ & $25\times 10^{-21}$ \\
DFM & A4B5G5 & $19\times 10^{-21}$ & $22\times 10^{-21}$ \\
\hline\hline
\end{tabular}
\end{center}
\end{table}

\subsection{Signal region}\label{sec:SignalRegion}

\begin{figure}[!t]
\includegraphics[angle=0,width=84mm]{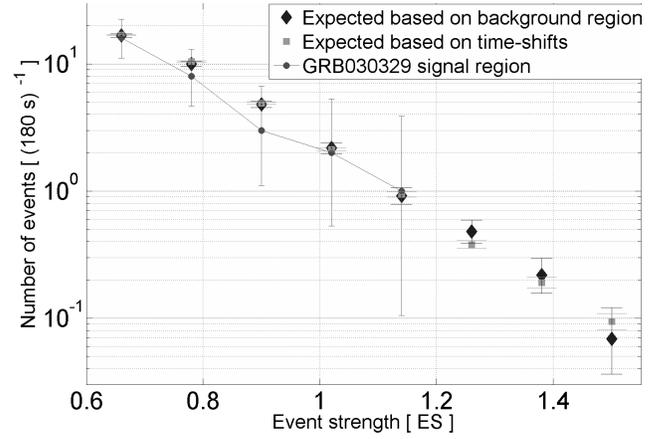}
\caption{Number of events versus event strength in the signal region
(circle). The diamonds show the expected distribution based on the
background region. The squares mark the expected distribution based
on non-physical time shifts (ranging from 2 to 9 seconds) between
the H1 and H2 datastreams in the background region. The error bars
reflect $90\%$ CL Poisson errors. The position of the symbols
correspond to the center of the bins.} \label{fig:SRH}
\end{figure}

The analysis of the signal region (Fig.~\ref{fig:SRH}) yielded
only events well below the predetermined event strength threshold
($\lesssim$60\% of threshold). Since we had no candidate event, we
placed an upper limit on the detected strength of gravitational
waves associated to GRB030329. Our fixed false alarm rate
permitted the results of simulations to be used directly in
setting upper limits.

The upper limits on $h_{RSS}$ for narrow band waveforms are given
in Table~\ref{t:NBET}. Tables~\ref{t:BBET}~and~\ref{t:DFMT} show
the upper limits for the broadband simulations and astrophysically
motivated waveforms, respectively.

\subsection{Errors}\label{sec:Errors}

The analysis method, the procedures used to determine the
efficiencies, and the non-stationary nature of the data, all
contribute to the uncertainty associated with the results.

The efficiency (versus $h_{RSS}$) values have an estimated
${\simeq}11\%$ uncertainty due to our limited knowledge of the
calibrated response of our detectors. This estimate also accounts
for the slight difference in calibrated response between the signal
region and background data used for the simulations.

An additional uncertainty arises from the non-stationarity of the
data. The results of the simulations exhibit a slight dependence on
the choice of the actual data segments (``base" data) used for the
injections. This dependence was characterized via simulations using
numerous different sub-segments of the background data. We repeated
the full efficiency estimation process several times for the same
waveform, while injecting into various base data stretches. The
variation in the measured upper limits indicated ${\simeq}10\%$
uncertainty due to the dependency of our upper limits on the base
data. This uncertainty shall also account for the statistical error
due to the finite number of simulations used.

We characterized the detection efficiencies for each waveform
considered via fits of sigmoid functions (see for example
Figure~\ref{fig:EFF_250_8.9}). The fits agree well with the data,
but small differences are occasionally observed in the $\gtrsim90\%$
efficiency region. We estimate that using these fits can
underestimate the 90\% limits by $\lesssim5\%$.

The uncertainties listed above are taken into account by
specifying a total 15\% uncertainty for each measurement in
Figure~\ref{fig:SENS} and in all Tables.

The false alarm rate associated with the results was also measured.
The false alarm rate limit is based on the measurement with zero lag
data plus the 90\% confidence Poisson error bars. We have checked
the assumption of Poisson background statistics by examining the
time intervals between consecutive triggers and the variance in
trigger counts for varying ES thresholds when the background sample
is divided into 50 equal-length intervals. Good agreement with the
Poisson expectation is observed. This choice provides a conservative
estimate of our associated (${\simeq }$5$\times$10$^{-4}$~Hz) false
alarm rate.

\subsection{Astrophysical interpretation}\label{sec:Interpretation}

GRB030329 has a well-determined redshift, therefore we can relate
our observed limits on strain to a measure of the total
gravitational wave energy emission. For a strain $h(t)$ at
distance $D_L$ from a source of gravitational radiation, the
associated power is proportional to $\dot{h}^2$ ($\dot{h}=dh/dt$),
though the proportionality constant will depend on the (unknown)
emission pattern of the source and the antenna pattern of the
detector (for the known source position, but unknown polarization
angle).

In general, it is not possible to relate our upper limit on the
strain from a particular waveform to a limit on the energy radiated
by the source, without assuming a model. Sources that radiate energy
$E_{\mathrm{GW}}$ might produce an arbitrarily small signal $h(t)$
in the detector, e.g., if the dynamics in the source were purely
axi-symmetric with the detector located on the axis. Nevertheless,
we can associate a strain $h(t)$ in the detector with some minimum
amount of gravitational-wave energy radiated by the source by
choosing an ``optimistic'' emission pattern, thereby obtaining a
measure of the minimum amount of energy that would need to be
radiated in order to obtain a detectable signal. We will show that
the progenitor of GRB030329 is not expected to have produced a
detectable signal.

We are interested in a ``plausible case scenario'' of gravitational
wave emission in order to obtain the minimum (plausible) amount of
gravitational-wave energy radiated that could be associated with a
detector signal $h(t)$.  We do not expect the gravitational waves to
be strongly beamed, and we expect that we are observing the GRB
progenitor along some preferred axis.  We take a model best case
scenario to be that of gravitational wave emission from a triaxial
ellipsoid rotating about the same axis as the GRB (i.e., the
direction to the Earth). If we assume quadrupolar gravitational wave
emission, the plus- and cross-polarization waveforms, emitted at a
polar angle $\theta$ from the axis of rotation to be:
\begin{eqnarray}
  h_+ &=& \frac12 (1 + \cos^2\theta)\, h_{+,0} \\
  h_\times &=& \cos\theta\, h_{\times,0}
\end{eqnarray}
where $h_{+,0}$ and $h_{\times,0}$ are two orthogonal waveforms
(e.g., a Sine-Gaussian and a Cosine-Gaussian), each containing the
same amount of radiative power.  That is, we assume that the same
amount of gravitational-wave energy is carried in the two
polarizations and that they are orthogonal:
\begin{equation}
  \int_{-\infty}^\infty \dot{h}_{+,0}^2 \, dt
  = \int_{-\infty}^\infty \dot{h}_{\times,0}^2 \, dt
\quad\textrm{and}\quad
  \int_{-\infty}^\infty
  \dot{h}_{+,0}
  \dot{h}_{\times,0} \, dt = 0.
\end{equation}
Thus, we would expect that the gravitational waves travelling
along the rotational axis (toward the Earth) would be circularly
polarized, and that the detector would receive the signal
\begin{equation}
  h = F_+ h_{+,0} + F_\times h_{\times,0}
\end{equation}
where $F_+$ and $F_\times$ represent the detector responses to the
polarization components $h_{+,0}$ and $h_{\times,0}$
\cite{Thorne-300}, and depend on the position of the source in the
sky and on a polarization angle.  The radiated energy from such a
system is calculated to be
\begin{equation}
  E_{\mathrm{GW}} = \frac{c^3}{16\pi G}\int dA \int_{-\infty}^\infty
  (\dot{h}^2_+ + \dot{h}^2_\times) dt
  = \frac{c^3}{5G}\frac{D_L^2}{\eta^2} \int_{-\infty}^\infty \dot{h}^2 dt
\end{equation}
where $\eta^2=F_+^2+F_\times^2$ (which depends only on the
position of the source on the sky) and where we are integrating
over a spherical shell around the source with radius $D_L$ (the
distance to the Earth).  Alternatively, using Parseval's identity,
we have
\begin{equation}
  E_{\mathrm{GW}} = \frac{8\pi^2c^3}{5G}\frac{D_L^2}{\eta^2}
  \int_0^\infty |f\tilde{h}|^2 df
\end{equation}
where
\begin{equation}
  \tilde{h}(f)=\int_{-\infty}^\infty h(t) e^{-2\pi ift}\, dt.
\end{equation}
Whereas optimal orientation gives $\eta=1$ for a source at zenith,
the position of GRB030329 was far from optimal.  The angle with
respect to zenith was $68^\circ$ and the azimuth with respect to
the $x$-arm was $45^\circ$, which yields $\eta=0.37$.

We now relate $E_{\mathrm{GW}}$ to the strain upper limits using
the specific waveforms used in the analysis.  For a Gaussian
waveform [see Eq.~(\ref{Gaussian})]:
\begin{equation}\label{Int5}
  E_{\mathrm{GW}} = \left(\frac{\sqrt{\pi}c^3}{10G}\right)
  \left(\frac{D_L^2 h_\circ^2}{\sigma}\right)
\end{equation}
and for a sine-Gaussian waveform [see Eq.~(\ref{SineGauss})]:
\begin{equation}
  E_{\mathrm{GW}} = \left(\frac{\sqrt{\pi}c^3}{20G}\right)
  \left(\frac{D_L^2 h_\circ^2}{\sigma}\right)
  ( 1 + 2Q^2 - e^{-Q^2} )
\end{equation}
where $Q=\omega_\circ\sigma=2\pi f_\circ\sigma$.  The relation
between $h_\circ$ and $h_{RSS}$ is given in
Eqs.~(\ref{GaussianhRSS}) and~(\ref{SGhRSS}).

We can relate the observed limit on $h_{RSS}$ to an equivalent mass
$M_{\mathrm{EQ}}$ which is converted to gravitational radiation with
$100\%$ efficiency, $E_{\mathrm{GW}}=M_{\mathrm{EQ}}c^2$, at a
luminosity distance $D_L\approx 800\,\textrm{Mpc}$. For
sine-Gaussian waveforms with $f_\circ=250\,\textrm{Hz}$ and $Q=8.9$,
$M_{\mathrm{EQ}}=1.9\times10^4\eta^{-2}M_\odot$. For Gaussian
waveforms with $\sigma=1\,\textrm{ms}$,
$M_{\mathrm{EQ}}=3.1\times10^4\eta^{-2}M_\odot$. However, we would
not expect that the gravitational-wave luminosity of the source
could exceed $\simeq c^5/G=2\times10^5M_\odot c^2$ per
second~\cite{misner73a}, so we would not expect an energy in
gravitational waves much more than $\simeq 2\times10^3~M_\odot c^2$
in the $\simeq10\,\textrm{ms}$ Sine-Gaussian waveform, or an energy
of much more than $\simeq 3\times10^4 M_\odot c^2$ in the maximum
duration ($150\,\textrm{ms}$) of the search; far below the limits on
$M_{\mathrm{EQ}}c^2$ that we find in this analysis. Present
theoretical expectations on the gravitational wave energy emitted
range from 10$^{-6}$ M$_\odot$ c$^2$ - 10$^{-4}$ M$_\odot$ c$^2$ to
10$^{-1}$ M$_\odot$ c$^2$ - M$_\odot$ c$^2$ for some of the most
optimistic models [see e.g.~\cite{Mosquera02, Ruffini01, Putten04,
Araya-Gochez03}]. Nonetheless, these scalings indicate how we can
probe well below these energetic limits with future analyses. For
example, assuming similar detector performance for an optimally
oriented trigger like GRB980425 (D$_L\approx$35~Mpc) the limit on
the equivalent mass would be $M_{\mathrm{EQ}}\approx$60~M$_\odot$
for the Gaussian waveforms mentioned above with
$\sigma=1\,\textrm{ms}$.

\section{Summary}\label{sec:Discussion}

\subsection{Comparison with previous searches for gravitational waves from GRBs}\label{sec:Comparison}

Our result is comparable to the best published results searching for
association between gravitational waves and GRBs~\cite{Astone2004},
however these studies differ in their most sensitive frequency.

Tricarico et al.~\cite{Tricarico01} used a single resonant mass
detector, AURIGA~\cite{Prodi98} , to look for an excess in
coincidences between the arrival times of GRBs in the BATSE 4B
catalog. They used two different methods. They searched for events
identified above a certain threshold in the gravitational wave data,
and also attempted to establish a statistical association between
GRBs and gravitational waves. No significant excess was found with
the former method. The latter used a variant of the correlation
based Finn-Mohanty-Romano (FMR) method~\cite{FMR99}. However,
instead of using the cross-correlation of two detectors, as proposed
in the FMR method, only the variance of the single detector output
was used. A sample of variances from times when there were no GRBs
was compared with a corresponding sample from data that spanned the
arrival times of the GRBs.  An upper limit on the source-averaged
gravitational wave signal root mean square value of
$1.5\times10^{-18}$ was found using 120 GRBs. This limit applies at
the AURIGA resonant frequencies of 913 and 931 Hz, which are very
far from the most sensitive frequency of the LIGO detectors
($\simeq$250~Hz). This work ~\cite{Tricarico01} was later extended
~\cite{Tricarico03}, which led to an improved upper limit.

The data analysis method employed in Modestino \&
Moleti~\cite{Modestino02} is another variant of the FMR method.
Instead of constructing off-source samples from data segments that
are far removed from the GRB trigger, the off-source samples are
constructed by introducing non-zero time shifts between the two
detector data streams and computing their cross-correlation. For
narrowband resonant mass detectors, the directional information of
a GRB cannot be exploited to discriminate against incorrect
relative timing since the signal in the output of the detector is
spread out by the detector response over time scales much larger
than the light travel time between the detectors.

Astone et al.~\cite{Astone99,Astone02} report on a search for a
statistical association between GRBs and gravitational waves using
data from the resonant mass detectors EXPLORER~\cite{Astone93} and
NAUTILUS~\cite{Astone97}. They report a Bayesian upper limit on
gravitational wave signal amplitudes of 1.2$\times$10$^{-18}$, at
95\% probability, when the maximum delay between the GRB and
gravitational wave is kept at 400 sec. The upper limit improves to
6.5$\times$10$^{-19}$ when the maximum delay is reduced to 5 sec.
However, the absence of directional and/or distance information for
most of these GRBs precluded accounting for source variations; the
gravitational wave signal amplitude was assumed to be the same for
all of the GRBs.

Astone et al.~\cite{Astone99B} report on the operation of the
resonant mass detectors EXPLORER during the closest ever gamma ray
burst (GRB980425) with known redshift and direction. At the time of
the burst, EXPLORER was taking data with close to optimal
orientation. GRB980425 was $\simeq$23 times closer to Earth than
GRB030329 giving a $\simeq$520 increase in energy sensitivity. Based
on their sensitivity and the loudest event within $\pm$5~minutes of
the GRB980425 trigger the authors quote a limit of
$\simeq$1600~M$_\odot$ for a simple model assuming isotropic
gravitational wave emission.

Recently, Astone et al.~\cite{Astone2004} executed a search aiming
to detect a statistical association between the GRBs detected by the
satellite experiments BATSE and BeppoSAX, and the EXPLORER and
NAUTILUS gravitational wave detectors. No association was uncovered.
Their upper limit is the lowest published result, which is based on
bar-detector gravitational wave data.

\subsection{Conclusion}\label{sec:Conclusion}

We have executed a cross-correlation-based search for possible
gravitational wave signatures around the GRB030329 trigger, which
occurred during the Second Science Run (S2) of the LIGO detectors.
We analyzed a 180 second signal region around the GRB and 4.5 hours
of background data, surrounding the signal region, corresponding to
a single coincident lock stretch. These data were sufficient to
characterize the background, scan the signal region and estimate our
efficiency. We used the same procedure, based on cross correlation,
for each of these studies. We evaluated the sensitivity of the
search to a large number of broad and narrow band waveforms.

We observed no candidates with gravitational wave signal strength
larger than a pre-determined threshold, therefore we set upper
limits on the associated gravitational wave strength at the
detectors. The present analysis covers the most sensitive frequency
range of the Hanford detectors, approximately from 80~Hz to 2048~Hz.
The frequency dependent sensitivity of our search was h$_{RSS}
{\simeq}$O(6$\times$10$^{-21}$)~Hz$^{-1/2}$.

The prospect for future searches is promising, as the sensitivity of
the instruments improves with further commissioning.

Once operating at target sensitivity, the detectors will be more
sensitive to strain than they were during S2 by factors of 10~-~100,
depending on frequency (see Figure~\ref{fig:LIGOEFF}.). This implies
an improvement of a factor of $\sim$1000 in sensitivity to E$_{GW}$,
since E$_{GW}$ scales like $\sim h_{RSS}^{2}$ (see for example
Eq.~\ref{Int5}).

Detection of GRBs with measured redshifts significantly smaller
than GRB030329's is certainly possible. GRB030329's
electromagnetic brightness was due to a favorable combination of
distance and our position in its beam. One year of observation
will incorporate hundreds of GRBs with LIGO data coverage and some
of these GRBs, even though
fainter~\cite{Woosley2004,Sazonov2004,Soderberg2004} than
GRB030329, could be significantly closer, as was 1998bw. We can
also hope for sources with more optimal direction and coincidence
between three or four observing interferometers.

\section{Acknowledgements}\label{sec:Acknowledgement}

The authors gratefully acknowledge the support of the United States
National Science Foundation for the construction and operation of
the LIGO Laboratory and the Particle Physics and Astronomy Research
Council of the United Kingdom, the Max-Planck-Society and the State
of Niedersachsen/Germany for support of the construction and
operation of the GEO600 detector. The authors also gratefully
acknowledge the support of the research by these agencies and by the
Australian Research Council, the Natural Sciences and Engineering
Research Council of Canada, the Council of Scientific and Industrial
Research of India, the Department of Science and Technology of
India, the Spanish Ministerio de Ciencia y Tecnologia, the John
Simon Guggenheim Foundation, the Leverhulme Trust, the David and
Lucile Packard Foundation, the Research Corporation, and the Alfred
P. Sloan Foundation. We are grateful to Scott Barthelmy and the GCN
network and Kevin Hurley and the IPN network for providing us with
near real time GRB triggers and to the Ulysses, Konus, SAX, and HETE
experiments, who detect and generate the events distributed by GCN
and IPN. This research has made use of data obtained from the HETE
science team via the website http://space.mit.edu/HETE/Bursts/Data.
HETE is an international mission of the NASA Explorer program, run
by the Massachusetts Institute of Technology.

\bibliographystyle{apsrev}
\bibliography{References}

\end{document}